\documentclass{article}
\usepackage[english]{babel}
\usepackage[letterpaper,top=2cm,bottom=2cm,left=3cm,right=3cm,marginparwidth=1.75cm]{geometry}

\usepackage{amsmath}
\usepackage{graphicx}
\usepackage{color}
\usepackage[colorlinks=true, allcolors=blue]{hyperref}
\usepackage{authblk}

\usepackage{ifthen} 
\newboolean{uprightparticles}
\setboolean{uprightparticles}{false} 

\usepackage{xspace} 
\usepackage{upgreek}

\def\atlas  {\mbox{ATLAS}\xspace}
\def\cms    {\mbox{CMS}\xspace}

\def\belle  {\mbox{Belle}\xspace}

\def\MagUp {\mbox{\em Mag\kern -0.05em Up}\xspace}

\ifthenelse{\boolean{uprightparticles}}%
{

 \def\PDelta      {\ensuremath{\Delta}\xspace}                 
 \def\PXi      {\ensuremath{\Xi}\xspace}                 
 \def\PLambda      {\ensuremath{\Lambda}\xspace}                 
 \def\PSigma      {\ensuremath{\Sigma}\xspace}                 
 \def\POmega      {\ensuremath{\Omega}\xspace}                 
 \def\PUpsilon      {\ensuremath{\Upsilon}\xspace}                 
 
 \def\PB      {\ensuremath{\mathrm{B}}\xspace}                 
                  
 \def\PD      {\ensuremath{\mathrm{D}}\xspace}

 \def\PK      {\ensuremath{\mathrm{K}}\xspace}

 \def\Pi      {\ensuremath{\mathrm{i}}\xspace}

}
{

 \mathchardef\PDelta="7101
 \mathchardef\PXi="7104
 \mathchardef\PLambda="7103
 \mathchardef\PSigma="7106
 \mathchardef\POmega="710A
 \mathchardef\PUpsilon="7107
                  
 \def\PB      {\ensuremath{B}\xspace}                 
                  
 \def\PD      {\ensuremath{D}\xspace}

 \def\PK      {\ensuremath{K}\xspace}

 \def\Pi      {\ensuremath{i}\xspace}

}

\makeatletter
\ifcase \@ptsize \relax
  \newcommand{\miniscule}{\@setfontsize\miniscule{4}{5}}
\or
  \newcommand{\miniscule}{\@setfontsize\miniscule{5}{6}}
\or
  \newcommand{\miniscule}{\@setfontsize\miniscule{5}{6}}
\fi
\makeatother

\DeclareRobustCommand{\optbar}[1]{\shortstack{{\miniscule (\rule[.5ex]{1.25em}{.18mm})}
  \\ [-.7ex] $#1$}}

  \def\Kbar    {{\kern 0.2em\overline{\kern -0.2em \PK}{}}\xspace}

\def\KorKbar    {\kern 0.18em\optbar{\kern -0.18em K}{}\xspace}

  \def\Dbar    {{\kern 0.2em\overline{\kern -0.2em \PD}{}}\xspace}

\def\DorDbar    {\kern 0.18em\optbar{\kern -0.18em D}{}\xspace}

\def\Bbar    {{\ensuremath{\kern 0.18em\overline{\kern -0.18em \PB}{}}}\xspace}

\def\BorBbar    {\kern 0.18em\optbar{\kern -0.18em B}{}\xspace}

  \def\Y#1S{\ensuremath{\PUpsilon{(#1S)}}\xspace}

\def\Lbar        {{\ensuremath{\kern 0.1em\overline{\kern -0.1em\PLambda}}}\xspace}
\def\LorLbar    {\kern 0.18em\optbar{\kern -0.18em \PLambda}{}\xspace}

\def\AT#1     {\ensuremath{A_{\mathrm{T}}^{#1}}\xspace}           

\def\C#1      {\ensuremath{\mathcal{C}_{#1}}\xspace}                       
\def\Cp#1     {\ensuremath{\mathcal{C}_{#1}^{'}}\xspace}                    
\def\Ceff#1   {\ensuremath{\mathcal{C}_{#1}^{\mathrm{(eff)}}}\xspace}        
\def\Cpeff#1  {\ensuremath{\mathcal{C}_{#1}^{'\mathrm{(eff)}}}\xspace}       
\def\Ope#1    {\ensuremath{\mathcal{O}_{#1}}\xspace}                       
\def\Opep#1   {\ensuremath{\mathcal{O}_{#1}^{'}}\xspace}                    

\newcommand{\tev}{\ifthenelse{\boolean{inbibliography}}{\ensuremath{~T\kern -0.05em eV}}{\ensuremath{\mathrm{\,Te\kern -0.1em V}}}\xspace}
\newcommand{\gev}{\ensuremath{\mathrm{\,Ge\kern -0.1em V}}\xspace}
\newcommand{\mev}{\ensuremath{\mathrm{\,Me\kern -0.1em V}}\xspace}
\newcommand{\kev}{\ensuremath{\mathrm{\,ke\kern -0.1em V}}\xspace}
\newcommand{\ev}{\ensuremath{\mathrm{\,e\kern -0.1em V}}\xspace}
\newcommand{\gevc}{\ensuremath{{\mathrm{\,Ge\kern -0.1em V\!/}c}}\xspace}
\newcommand{\mevc}{\ensuremath{{\mathrm{\,Me\kern -0.1em V\!/}c}}\xspace}
\newcommand{\gevcc}{\ensuremath{{\mathrm{\,Ge\kern -0.1em V\!/}c^2}}\xspace}
\newcommand{\gevgevcccc}{\ensuremath{{\mathrm{\,Ge\kern -0.1em V^2\!/}c^4}}\xspace}
\newcommand{\mevcc}{\ensuremath{{\mathrm{\,Me\kern -0.1em V\!/}c^2}}\xspace}

\def\gsim{{~\raise.15em\hbox{$>$}\kern-.85em
          \lower.35em\hbox{$\sim$}~}\xspace}
\def\lsim{{~\raise.15em\hbox{$<$}\kern-.85em
          \lower.35em\hbox{$\sim$}~}\xspace}

\def\dirac      {\mbox{\textsc{Dirac}}\xspace}

\def\root       {\mbox{\textsc{Root}}\xspace}

\def\cpp        {\mbox{\textsc{C\raisebox{0.1em}{{\footnotesize{++}}}}}\xspace}

\def\tell1  {TELL1\xspace}
\def\ukl1   {UKL1\xspace}

\usepackage{xspace}

\def\cvmfs{{\tt CVMFS}\xspace}
\def\eos{{\tt EOS}\xspace}
\def\root{{\tt ROOT}\xspace}
\def\posix{{\tt POSIX}\xspace}
\def\json{{\tt JSON}\xspace}
\def\git{{\tt Git}\xspace}
\def\cpp{{\tt C++}\xspace}
\def\python{{\tt Python}\xspace}
\def\twiki{{\tt Twiki}\xspace}
\def\lhcbdirac{{\tt LHCbDIRAC}\xspace}
\def\dirac{{\tt DIRAC}\xspace}

\title{\Huge{Constraints on future analysis metadata systems
in High Energy Physics}}

\author[5]{T. J. Khoo} 
\author[10]{A. Reinsvold Hall} 
\author[16]{N. Skidmore} 
\author[15]{S. Alderweireldt} 
\author[13]{J. Anders} 
\author[3]{C. Burr} 
\author[9]{W. Buttinger} 
\author[11]{P. David} 
\author[3]{L. Gouskos} 
\author[4]{L. Gray} 
\author[3]{S. Hageb\"ock} 
\author[3]{A. Krasznahorkay} 
\author[1]{P. Laycock} 
\author[14]{A. Lister} 
\author[6]{Z. Marshall} 
\author[2]{A. B. Meyer} 
\author[2]{T. Novak} 
\author[12]{S. Rappoccio} 
\author[7]{M. Ritter} 
\author[8]{E. Rodrigues} 
\author[3]{J. Rumsevicius} 
\author[4]{L. Sexton-Kennedy} 
\author[4]{N. Smith} 
\author[3]{G. A. Stewart} 
\author[11]{S. Wertz} 

\affil[1]{\it Brookhaven National Laboratory Physics Department, Upton, NY, US}
\affil[2]{\it Deutsches Elektronen-Synchrotron DESY, Germany}
\affil[3]{\it European Organization for Nuclear Research (CERN), Geneva, Switzerland}
\affil[4]{\it Fermi National Accelerator Laboratory, Batavia, IL, USA}
\affil[5]{\it Humboldt-Universität zu Berlin, Institut für Physik, 12489 Berlin, DE}
\affil[6]{\it Physics Division, Lawrence Berkeley National Laboratory, Berkeley, CA, USA}
\affil[7]{\it Ludwig-Maximilians-Universit\"at, M\"unchen, DE}
\affil[8]{\it Oliver Lodge Laboratory, University of Liverpool, Liverpool, UK}
\affil[9]{\it Rutherford Appleton Laboratory, Didcot, OX11 0DE, UK}
\affil[10]{\it United States Naval Academy, Annapolis, MD, USA}
\affil[11]{\it Université Catholique de Louvain, Centre for Cosmology, Particle
Physics and Phenomenology (CP3), Louvain-la-Neuve, Walloon Brabant, Belgium}
\affil[12]{\it University at Buffalo, State University of New York, Amherst, NY, USA}
\affil[13]{\it University of Bern, Laboratory for High Energy Physics, Bern, CH}
\affil[14]{\it University of British Columbia, Vancouver, BC, CA}
\affil[15]{\it The University of Edinburgh, Edinburgh, GB}
\affil[16]{\it University of Manchester, Schuster Building, Manchester, M13 9PL, UK}

\date{\today}

\begin{document}

\maketitle

\begin{abstract}
In High Energy Physics (HEP), analysis metadata comes in many forms -- from theoretical cross-sections, to calibration corrections, to details about file processing. Correctly applying metadata is a crucial and often time-consuming step in an analysis, but designing analysis metadata systems has historically received little direct attention. Among other considerations, an ideal metadata tool should be easy to use by new analysers, should scale to large data volumes and diverse processing paradigms, and should enable future analysis reinterpretation. This document, which is the product of community discussions organised by the HEP Software Foundation, categorises types of metadata by scope and format and gives examples of current metadata solutions. Important design considerations for metadata systems, including sociological factors, analysis preservation efforts, and technical factors, are discussed. A list of best practices and technical requirements for future analysis metadata systems is presented. These best practices could guide the development of a future cross-experimental effort for analysis metadata tools.

\end{abstract}


\section{Introduction}
\label{sec:intro}

This document attempts to motivate and codify a set of requirements on systems for storing, organising and accessing analysis metadata (henceforth ``metadata systems'') for future experiments in High Energy Physics (HEP). It arises from a series of meetings held by the Data Analysis Working Group (DAWG) of the HEP Software Foundation (HSF)\footnote{https://hepsoftwarefoundation.org/} in early 2021 and the subsequent cross-experiment discussions.

High Energy Physics is a big data endeavour, and significant research efforts have been dedicated to the management of and access to physics data, which is usually interpreted as the combination of detector readout and simulated events. While metadata (\mbox{i.e.} data about data) has long been an important part of detector simulation and reconstruction, in the context of analysis it has received less direct attention. Metadata is a crucial element of data analysis and analysis preservation, but in the software for current experiments, such as those at the Large Hadron Collider (LHC), tools for handling it have less frequently been designed with analysis applications in mind. Instead, analysis metadata systems were either adapted from systems designed for central dataset production or emerged ad-hoc from within the analysis community. In future HEP endeavours, such as the LHC’s High Luminosity upgrade (HL-LHC)~\cite{hllhc}, the lack of a coordinated approach to storing and retrieving analysis metadata may become a limiting factor in the efficiency of analysis. The longer lifetimes of future HEP experiments mean that ever larger datasets, spanning many data-taking periods with changing conditions, have to be analysed. Consequently, analyses are bigger, more collaborative enterprises that will require more coherent, persistent metadata solutions.

In this document, we will first give examples of different types of metadata and their various scopes. Then we will discuss motivations for the design choices of metadata systems, including current examples. Finally, we will outline technical specifications that should be considered during the design of any future metadata system and identify how these fulfill the ``FAIR Guiding Principles for scientific data management and stewardship" \cite{FAIR}.   As the issues motivating these are common to many HEP experiments, these could become the foundation of a common cross-experimental software project, or could help define the specifications of systems developed within the experimental collaborations.

\subsection*{Types of Metadata}

For the purposes of this discussion, metadata is taken to refer to any information other than the event data of a simulated or recorded dataset. For example, in the context of an LHC experiment, a recorded bunch crossing is described by event data in the form of tracker hits, calorimeter cell energies and other detector readout, while the corresponding metadata may include the instantaneous luminosity, magnetic field conditions, data quality assessments and so on. These need not be stored in the same location. Frequently, the event data is streamed to \root ~\cite{ROOT} files containing tree data structures, with some metadata retained in the file, but other details being relegated to a relational database (henceforth “database”), using identifiers to associate database information to specific files or datasets. Still more relevant information needed to correctly analyse the events may be stored in other formats.

An extensive, but by no means exhaustive, list of metadata examples is given below, together with some mechanisms by which the metadata is stored and accessed. All of these are found to be necessary for carrying out LHC data analysis. In this document, we focus on the information needed to carry out the analysis, rather than on information about analysis team members or other collaboration details. Important as this is for collaboration organisation, it does not impact computing requirements as strongly and does not impact the reproducibility of the results.

\begin{enumerate}
    \item Dataset provenance -- software versions used to generate or process the data, input and affiliated event samples. The full information is typically stored in databases, but some information is available in files or encoded in dataset names.
    \item Book-keeping information -- cut flow records from filtering samples during processing, as well as initial (possibly weighted) numbers of events generated, which are important for normalising MC generated samples. These values are typically stored in files, potentially as a dedicated data structure or, \mbox{e.g.}, as \root histograms filled during processing.
    \item Data quality assessments -- flags or lists indicating whether blocks of data are suitable for analyses, at varying levels of granularity, which may also be used for luminosity calculations. Event-level information may be recorded as flags in the data files, while other data formats, \mbox{e.g.} databases or {\tt XML} files, may be used to describe data quality assessments over longer timescales.
    \item Calibration data  -- incredibly diverse, from detailed detector information including alignment constants and magnetic field conditions to calibration corrections for physics object selection efficiencies or four-momenta. Consequently, this type of metadata tends to be stored in a myriad of forms such as databases, text files or code in version-control repositories, \root files in common \eos~\cite{eos} or \cvmfs~\cite{cvmfs} directories, and text or attachments on webpages.
    \item A special case of calibration data is that of reweighting information, used to correct distributions across the full dataset, and potentially requiring recalculation for different samples. Common applications include correcting the distribution of the generated pileup multiplicity or adjustments of event kinematics based on control regions or published measurements.
    \item Information pertaining to Monte Carlo (MC) datasets -- notably event generator input parameters, which are often not published in full detail, and production cross-sections, which in complex signal samples may need to be correlated with the subprocess generated in every event. This information may be stored in a database, but commonly needs to be looked up from tables on \twiki pages\footnote{For example, the webpage of the LHC Higgs Working Group (https://twiki.cern.ch/twiki/bin/view/LHCPhysics/LHCHWG) or the LHC SUSY Cross-Section Working Group (https://twiki.cern.ch/twiki/bin/view/LHCPhysics/SUSYCrossSections)} or other common filesystem storage.
    \item An emerging feature is the ability to add user tags to datasets for dataset discovery, organisation of production campaigns, or other purposes. This is currently being done in MC production databases, for example in the ATLAS Metadata Interface (AMI) system~\cite{AtlasMD}.

\end{enumerate}

\subsection*{Metadata scopes}

The examples above may be loosely classified by the scope of the data to which they pertain:

\begin{itemize}
\item Analysis metadata (including examples 4 and 7 above) -- describes features of an analysis, such as lists of required datasets and how they are used, versions of calibration metadata used to produce final results, and so on. “Datasets” here refer to samples of events as they are organised in persistent storage, usually according to some useful common criteria, \mbox{e.g.}, data recorded during the same run with the same triggers or data simulated with common parameters
\item Dataset metadata (includes examples 1, 6, and 7, arguably 4 and 5) -- describes either features of datasets, or information about how to analyse datasets. 
\item Time-dependent metadata (includes examples 3, 4, and 5) -- describes information that varies over the course of data collection, typically by being tied to timestamps on the detector data, defining “intervals of validity” (IOVs). In the case of calibration data derived through analysis of simulated or recorded data, IOVs may be as wide as “one specified year” or “one multi-year run”, and handled similarly to dataset metadata.
\item File-dependent metadata (includes examples 1, 2, and 3) -- information about a single file, therefore typically related to the mechanics of file processing. Note that this is not the same as metadata stored in the file, which may in fact be dataset metadata or time-dependent metadata.
\end{itemize}

\section{Motivations}
\label{sec:motivations}

\subsection*{Sociological factors}

One of the major challenges when designing any metadata system is encouraging widespread use of the tool(s). There are many examples of useful, well-intentioned tools that failed to be adopted by the community and were eventually replaced. A key aspect that was discussed during the analysis metadata workshops is a good user interface. Analysers prefer a \posix -like command-line/scriptable interface 
over web access to a database, with the overhead of repeated authentication (\mbox{e.g.} via the {\tt X.509} protocol\footnote{https://www.itu.int/rec/T-REC-X.509}) being considered one of the disadvantages of the latter. Reading information off \cvmfs is one popular \posix -like approach that can accommodate unobtrusive authentication procedures. Any new metadata system needs to give careful consideration to how to incentivise analysers to use the system as intended.

Another challenge here is coordinating who is allowed to update the information stored in the metadata system and ensuring that there is sufficient personpower to keep any metadata system up-to-date with the latest recommendations. In some cases, such as information extracted after dataset production (\mbox{e.g.} N(N)LO cross-sections or k-factors), it makes sense to allow for vetted user submissions.  In other cases, such as centrally derived corrections to the final physics four-momenta, it makes sense to restrict write-access to specific people or groups. The metadata tools should be flexible enough to handle both situations. If users do not feel they can trust the results in a given system, then that encourages more ad-hoc solutions such as looking up corrections on various \twiki pages. Overcoming the concept that “busywork equals validation” may require effective and convenient validation tools and simple APIs to update the metadata as needed. Training is important, especially for new analysers who do not know who to ask.

\subsection*{Analysis preservation}

Metadata systems should be designed to satisfy the requirements of analysis preservation, \mbox{i.e.} the capability to repeat an analysis workflow for reproduction or reinterpretation after the analysis has been finalised, a la {\tt RECAST}\footnote{https://github.com/recast-hep} or {\tt REANA}\footnote{https://reanahub.io/} and other similar projects. Different requirements may apply depending on whether the reproduction is meant to be collaboration-internal, with access to the full software stack, or publicly executable. In either case, the analysis description needs to be recorded in sufficient detail so that it can be reimplemented. In the case that this is addressed via analysis code preservation, then event data and metadata dependencies must be preserved as well, together with access APIs.

For many current analyses, the key metadata sources --- \mbox{e.g.}, which corrections were applied or even what datasets were used --- are only documented in internal notes or \twiki pages. This makes it difficult or impossible for future analyzers or theorists to accurately reproduce the results. Capturing all of the analysis metadata associated with a published result is therefore an important goal when designing an analysis system. For analysis preservation, clear versioning is also crucial. But there can be significant complexity involved, as analysis groups often wish to use different versions for individual corrections. In order to validate corrections and calibrations and compare between different sets of conditions, good tools to customise and inspect metadata payloads are a necessity.

There are several promising solutions in existing metadata systems. For example, Belle II has a well-tested infrastructure for analysis-conditions handling that relies on metadata “global tags”, in other words single identifiers encapsulating the full metadata configuration~\cite{belletwo}. This simplifies the documentation of metadata used by an analysis, because the relevant information is encoded in the global tags. Multiple global tags can be passed to the software framework configuration, and a framework service takes care of presenting this as though it were one global tag. Each analysis group in Belle II can use this to define their own lightweight “analysis global tags” that only need to capture conditions that are not included in the centrally managed reconstruction and simulation global tags. Thus while analysis preservation requires recording the complete set of global tags that were used for an analysis (usually a handful), the analysis global tags are only as complicated as they need to be. 

LHCb offline data processing steps up to creation of analyst level datasets are centrally preserved in the \lhcbdirac\footnote{https://lhcb-dirac.readthedocs.io/} book-keeping. \lhcbdirac is an extension of the \dirac Grid solution~\cite{dirac} that implements the LHCb data processing workflows, now including the creation of analysis specific datasets containing custom high-level physics variables through the use of Analysis Productions\footnote{LHCb DPA project, https://indico.desy.de/event/28202/contributions/105606/}. Analysis Productions are submitted by individual analysts declaratively via {\tt YAML} files by providing the job configuration and input data. To provide assurance that user-prepared configurations are correct, extensive tests are run on the {\tt GitLab}\footnote{https://about.gitlab.com/} Continuous Integration platform prior to approval. The \lhcbdirac bookkeeping system preserves metadata such as detector and data-taking conditions used to process the data, versions of applications used and the corresponding options files, enabling high-quality analysis preservation.

Another important consideration is the stability and longevity of the metadata formats throughout and beyond the lives of the experiments. That is, starting with a sound design should permit stability, but the strategy also needs to be adaptable enough to support changes in behaviour and available resources. For example, \cms is currently improving their approach by pushing for a unified \json format for metadata, with files stored in a central \git repository\footnote{https://git-scm.com/}. These choices ensure longevity of the payload and flexibility to accommodate diverse needs and take advantage of a widely used versioning system.

\subsection*{Book-keeping}

As recorded datasets grow, so does the lead time for investigative operations on these and associated simulated event samples. A recurring challenge in distributed analysis is handling small fractions of job failures due to intermittent technical failures at computing sites. Fully processing the last few percent of events may take several times longer than the bulk, leading to significant lost productivity. For final results, processing 100\% of the recorded events is necessary, whereas this criterion can be relaxed for intermediate studies, provided the capability exists to correctly scale results to the full target integrated luminosity.

In analysis of simulated datasets, the main requirement is that the generated number of events (sum of weights in the case of weighted datasets) for the processed files is accessible at the stage when analysis yields are determined. Additional attention may be needed in certain cases, \mbox{e.g.} where procedures involving reweighting by ratios of data and simulated distributions require knowing the full simulated distribution. When sample sizes are limited, variations or biases in the available samples may lead to undesirable effects.

\section{Technological considerations}
\label{sec:tech-considerations}

A future analysis metadata system must meet a number of technological criteria to function. The sociological and analysis preservation factors already discussed must be accommodated in its design, as there is a strong precedent of workarounds being established to achieve perceived goals of simplicity, even at the cost of robustness.

\subsection*{Metadata formats}

It is likely impossible to use a single format for all analysis metadata, but the identification of a few specific formats that effectively and flexibly accommodate the chief use cases is an important consideration in metadata system design. If specifications are absent or insufficiently versatile to meet the needs of metadata providers, formats may proliferate, complicating the infrastructure needed to serve the metadata, and increasing the burden on users. This situation emerged in the experience of ATLAS and CMS in the first two LHC runs, where analysis calibrations derived under time pressure were encoded in a wide variety of formats, with unified repositories invented only after the fact. Specific formats addressing the aforementioned metadata scopes (dataset, file, and time-dependent) are discussed below.

In-file metadata is natural for information that may be needed at the point of job configuration, avoiding the overhead and connectivity requirements of database lookup. It has been pointed out that the boundary between in-file metadata and event data is essentially arbitrary, particularly in systems that have more fluid hierarchical levels than {\tt ROOT}'s {\tt TTree}, permitting better optimisation of metadata content that may span a few files vs. a few events vs. a full dataset.

Similarly, book-keeping of filter fractions and cut flow information from selections applied over multiple data processing steps is probably natural to keep local to the processed data. There may be advantages to allowing easy extension of cut flows by user code, at least if lightweight common libraries are available. Information must be recorded in some format to permit processing of fractional datasets, when the data volumes become large enough to preclude regularly analysing the full datasets prior to the final publication.

Databases are a natural repository for centrally defined dataset-scope information such as dataset provenance and production configurations, as well as user-added information including cross-sections and other contextual labels. Detector calibration metadata tends to be in databases as a standard, and could also accommodate object calibration information, particularly those calibrations that are derived frequently and applied or updated at analysis time. For high-throughput analysis jobs, access overheads need to be kept down, perhaps via network speed or processing improvements such as a query mechanism with robust caching. This could encourage the use of unified object calibration file formats, and a good user submission and validation interface would keep submission efficient.

There are several potential barriers to widespread database usage. One such barrier may be the need for authentication and the primacy of web-based tools for browsing the database, where analysers prefer \posix-like access with minimal (\mbox{e.g.} {\tt Kerberos}\footnote{https://web.mit.edu/kerberos/}) authentication. This could be overcome by providing simple APIs for payload retrieval, possibly including export of a version of relevant metadata to a local or distributed disk location \mbox{e.g.} \cvmfs. Effective web interfaces are also needed to support browsing and comparison operations on versions of interest. \git hosting sites are one concrete example permitting version diffs, which are widely used.

Another barrier, particularly for smaller collaborations, is the expertise, person power, and resources needed to operate a database at scale. One tested solution (described in more detail below) to this problem is to factorise the database that captures the metadata versions and dependencies from the actual metadata payloads, which can be offloaded to a file system or \git or elsewhere. All payloads may not need to be stored in the same system, and in fact, the wide variety of metadata types that need to be accommodated may encourage heterogeneous systems rather than a monolith. A consistent payload location system, however, is desirable. Finally, any metadata database needs to be able to swiftly integrate with the actual scripts, notebooks, and other tools used by analysers. These last two points are discussed in more detail below.

\subsection*{Repository structure}

Some form of versioning is strictly necessary. First and foremost, analysers need to be able to specify which version of the metadata is to be used, whether for validation of changes that have been made, or to ensure reproducibility once analysis design has been “frozen” (\mbox{e.g.}, post-unblinding). Examples of versioning methods include:

\begin{itemize}
    \item As mentioned previously, \belle II uses the “global tag” formalism, adopted from conditions databases, adding the capability to merge tags in order to override configurations for specific subsets of the metadata.
    \item \atlas and lately \cms uses a set of write-once directories provided on {\tt CVMFS}, holding analysis calibration data. Write protection and timestamps or version numbers as directory paths constitute a free-form versioning solution that minimises constraints on developers.
    \item \atlas analysis recommendations are tightly tied to analysis software release tags of the collaboration’s common “Athena” software repository on {\tt Gitlab},
    wherein the \cpp or \python code includes default or hard-coded paths to the calibration data held on \cvmfs. Overrides of these defaults in analysis configurations may themselves be committed to \git together with the analysis frameworks.
    \item The new \json format for \cms calibration metadata will be self-documenting and include the version number directly in the file.
    \item For detector and data-taking conditions, LHCb uses {\tt GitCondDB}~\cite{gitconddb}, an experiment-independent conditions database system that leverages \git to manage versions and tags. The conditions database uses a 3-dimensional structure dependent on the condition IDs in question (\mbox{e.g.}, an {\tt XML} file), the version (a tag or branch name)
    and the IOV. A bare clone of the conditions repositories is distributed on \cvmfs making the conditions data available to all jobs, both local and distributed.
\end{itemize}

Transparency and analysis preservation are likely served well by the capability to define the full metadata versioning information in a single identifier. The metadata identifier should encapsulate the complete description of metadata used for an analysis, but as mentioned previously, the identifier could still point to \git commits in a repository, specific files on \cvmfs, or other locations where the metadata payloads are stored. In-file metadata, in particular metadata related to the details of file processing, are a likely exception to this goal. For flexibility, a scheme by which such identifiers can be customised and combined --- overriding parts of a generic “tag” with more specific requirements --- would be required.

There is a balance to be struck between convenience, \mbox{e.g.} allowing the latest version of any/all metadata to be used if not explicitly overridden during R\&D periods, and stability, \mbox{i.e.} avoiding silent changes under the feet of analysers. Choosing or even defining new tags for analysis milestones or publication addresses the latter requirement.
The point arose that explicit instructions to update settings are considered a way to prevent such unexpected changes, but this effectively equates to busywork that could be avoided with a more robust validation system. Rather, schemes for transparently associating metadata settings and versions with the resulting analysis datasets should be investigated. These would improve reproducibility and facilitate debugging of unexpected issues as needed.

The metadata content should be able to be served from multiple locations. For efficient use of resources, a subset of the metadata should be extractable based on user configuration. This is needed for distributed analysis at sites that may have limited connectivity, such as High Performance Computing and cloud computing nodes, as well as for local analysis on user laptops while in transit or otherwise away from a fast internet connection. Enabling analysis for people who may not have a fast connection to the main processing sites is also an important goal in the interest of equity, diversity, and inclusion. Some existing solutions for relocatability include:

\begin{itemize}
    \item Runtime download of required files to a local cache via {\tt http},  implemented as an option in the \atlas “PathResolver” code, which serves as a file search interface with awareness of a variety of sources including \git and the \cvmfs calibration area.
    \item On Belle II, the payload data are referenced in the relational database as a {\tt URL}, which allows the payload server information to be prepended to the filepath. Much like the \atlas PathResolver, the client can then specify alternative sources (\cvmfs is the most common) and a failover strategy (local, \cvmfs, central server).  The central server is the main repository and payloads are copied to \cvmfs after a short delay, meaning that the majority of read cases are supported by \cvmfs.  It is also possible to specify a local (squid) cache by setting a proxy at a computing site, which then takes care of pulling the information only once from the central server, which is only usually necessary if files are not available on \cvmfs for some reason (\mbox{e.g.} prompt calibration).
\end{itemize}

User submissions can enrich metadata and avoid duplication of the work involved in extracting commonly used information such as higher order cross-sections or k-factors. On ATLAS, this information has been migrated into the ATLAS Metadata Interface, where a restricted group of shifters reporting to the ATLAS Physics Modelling Group is permitted to upload new values requested via issue tickets, tagged with a timestamp.
Validation of such submissions is crucial. Consequently, the submission procedure should incorporate steps for checking correctness of the submitted information. The specifics of such checks are inextricable from domain expertise, but at bare minimum, the capability to perform value comparisons and syntax or format checking upon submission would reduce the risk of the most basic errors.

Another special case of user-added information is labelling of datasets with context beyond the mechanics of their production. This includes association of datasets with specific analyses or applications (background, signal, systematic variation, etc.). For example, LHCb has recently developed such functionality whereby datasets are automatically tagged by properties such as data-taking year and magnet polarity but additionally can be given custom tags to identify them for use in a particular analysis or shared among several. This aids analysis preservation, provides a straightforward and safe way to share analysis datasets, and prevents the need for hard-coded file paths.

While there is a mild risk that such labels could proliferate almost infinitely, there is significant potential to improve analysis workflows by simplifying dataset discovery and identification. These labels could also help to improve data curation operations: automating the obsolescence and deletion of unused data, or notifying analysers automatically when a problem is identified with a dataset. This application motivates interfacing dataset metadata tools with dataset management infrastructure and job management systems. This functionality is now supported in the Rucio~\cite{rucio} dataset management system.

\subsection*{Access interface}

As the diversity of analysis metadata storage systems would suggest, there is no uniform approach to providing APIs for accessing analysis metadata. There is something of a philosophical split, where communities such as ATLAS that favour centralising analysis metadata  have invested correspondingly in a uniform access layer not only for extracting information from metadata stores, but also transforming this information into a form directly applicable to analysis. This may inject a higher dependence on the broader collaboration software stack than is needed for direct metadata access. Other communities have favoured a minimalist style, avoiding these sorts of dependencies in pursuit of lightweight analysis code with fewer restrictions on users. However, this may lead to a larger implementation and validation burden on users.

In future metadata systems, a simple API should be provided for requesting and retrieving metadata payloads. This is important for ensuring frictionless access in analysis code. Ideally, access tools should have minimal dependencies and be easily installable, so as to minimise restrictions on analysis framework design and analysis job payload sizes, rather than coupled to heavier collaboration software libraries that may be more cumbersome to install and use on local hardware. The ability to preload only required metadata based on job configuration may be helpful for optimising metadata access for distributed analysis jobs. A few issues need to be taken into account in considering these targets.

Providing raw metadata payloads through a minimal interface may be straightforward via REST-ful APIs~\cite{Fielding2000}. However, depending on how the payloads themselves are structured, it may be more or less convenient to translate them into a form that is useful for analysis. For example, it may be efficient to store calibration information in the form of parameters for fitted functions, which if sufficiently complex may be impractical for users to reimplement. While this need not directly impact the metadata system itself, a choice may need to be made between defining a more substantial adjacent software layer for application of the stored metadata, and choosing a simpler format that may be less precise or efficient to apply.

Trends in analysis software evolution may also have implications for the programming model by which the metadata access tools are provided. In the near term, both \cpp and \python are likely to make up a significant share of analysis codes, with the adoption of data science tools and columnar analysis models growing. Consequently, multilingual support may be needed (in the absence of strong collaboration enforcement of language choices). Distribution via package installers or managers may be desirable.

The development of new analysis formats like {\tt NanoAOD} and {\tt DAOD\_PHYSLITE} ~\cite{nanoaod, atlasrunthree} may affect the degree of metadata access needed by the end user. On one hand, routine calibration operations might be applied centrally rather than in user code. On the other, file size reduction may require some operations, notably the application of systematic variations, to be done in memory.
\section{Technical specification}
\label{sec:tech-specification}

Taking into account the motivations discussed in Section \ref{sec:motivations} and the technological considerations from Section \ref{sec:tech-considerations}, we specify a general set of requirements and desired features for any future analysis metadata systems in HEP or in experiments facing similar challenges. Here the focus is on features that a future system needs to satisfy, rather than an attempt at prescribing a specific solution. See the previous sections for examples of how current metadata systems satisfy these requirements. 

\subsection*{Technical requirements}

\begin{itemize}
    \item Versioning 
    \begin{itemize}
        \item Reproducibility of analysis results---and principle F1 of the ``Findable" guiding principle of FAIR---demand that metadata identifiers access payloads that are immutable once published. 
        Payloads themselves will need to be periodically updated, in some cases frequently, without invalidating earlier versions. Therefore the payloads will need to be hosted on systems that support a write-once model with version tracking. 
        \item Versioned payloads of different types will need to be combined in order to serve the needs of a full analysis, for example particle ID or calibration recommendations for a wide range of objects. A mechanism for aggregating payload versions will be required to cleanly communicate these groupings. User friendly interfaces for browsing and comparing single or aggregate identifiers will be needed, with both scriptable and web APIs likely being important. The ability to register and search for relevant metadata is also part of the ``Findable" guiding principle of FAIR (F4).
        \item For experimentation and incremental changes, combinations of aggregate identifiers may be needed. Merging and override capabilities for the combination of identifiers will be needed, and the syntax should permit no ambiguity in these operations.
    \end{itemize}
    \item Relocatability 
    \begin{itemize}
        \item Distribution of the metadata payloads to multiple sites is needed in order to avoid connectivity bottlenecks and barriers. In particular, to ensure that workers without direct access to the hosting servers (\mbox{e.g.} at HPC facilities or for work while in transit), the capability to serve metadata from a local cache is needed. Partial caching of predetermined payloads is important for efficiency.
        \item Database lookups must be easy to redirect to a preferred source. 
    \end{itemize}
    \item Lightweight API 
    \begin{itemize}
        \item Diverse end-stage analysis code (\mbox{e.g.} \python or \cpp analysis frameworks) must be able to access the metadata, so access APIs should be lightweight with minimal dependencies. Programmatic access, \mbox{e.g.} in \python scripts, must be supported as a first-class use case. This ``interoperability" with standard analysis workflows is a FAIR guiding principle (I1).
        \item Protections should be implemented such that API usage cannot overload a database, even if a high volume of requests are made from batch or grid jobs. Both payload storage and identification systems need to have high availability and robustness, as well as aggressive caching of requests.
    \end{itemize}
    \item Extensible system 
    \begin{itemize}
        \item User submission of metadata must be allowed, to support various types of additions. The main use case will be for updates of metadata derived externally from central sample production systems, \mbox{e.g.} calibrations, cross-sections etc. A special case is the extension of cutflow information, which may be stored as in-file metadata.
        \item Robust access control systems will be needed to restrict additions to vetted submitters. Validation systems are also necessary, to ensure submissions can be tested stringently before being made accessible. 
        \end{itemize}
    \item Unobtrusive authentication 
    \begin{itemize}
        \item In relation to the ``Accessible" FAIR guiding principle (A1.2), access to the metadata content should be granted based on persistent authentication methods, rather than burdening users with repeated sign-in steps. This is particularly important for programmatic access to metadata.
    \end{itemize}
    \item Intervals of Validity
    \begin{itemize}
        \item The metadata identification system must have the ability to store information about relevant IOVs, particularly for partial-dataset calibrations. 
        \item This feature might also be usable for additional contextual configuration, such as identifying MC vs. recorded data. 
    \end{itemize}

\end{itemize}

\subsection*{Desirable features and other considerations}

\begin{itemize}
    \item Complete analysis description
    \begin{itemize}
        \item If all metadata descriptions can be captured in a single system, then it is natural to extend aggregation of metadata tags to provide a full description of software and inputs needed for a single analysis.
        \item Recording software versions, job configuration parameters, input datasets and auxiliary data fully serves multiple needs, including the tracking, preservation, or combination of analyses.
        \item While all analysis metadata identifiers would need to be encoded in the system, payloads themselves could be stored in any repository, although long-term analysis preservation requires guarantees on the longevity of all relevant storage systems.
    \end{itemize}
    \item User-applied dataset tags
    \begin{itemize}
        \item Free-form labels attached to event datasets can be used to serve various purposes, including identification and categorisation of analysis inputs, management of storage system capacity, and invalidation of obsolete or incorrect data. Allowing users to attach arbitrary labels and share those labels would allow substantial flexibility and would help the system satisfy the ``Reusable" FAIR guiding principle (R1).
        \item Directly interfacing the dataset labelling system to dataset management and job submission systems would generate additional hooks for efficient analysis management. 
    \end{itemize}
    \item Interpretation of metadata content
    \begin{itemize}
        \item Additional tooling, separate from the metadata access systems, may be useful for payload interpretation. For example, it may be efficient to store calibration corrections in the form of fit constants, or even as neural network parameters. Centrally provided tools would then be a more reliable solution for consistent and correct translation of the payloads to the final corrections.
    \end{itemize}
\end{itemize}

\section{Summary}
\label{sec:summary}

Based on the common experiences of analysers of LHC and contemporary particle physics experiments, this document has categorised types of analysis metadata in terms of scope and content, as well as identifying not only current technological solutions but also major challenges in the storage and access of analysis metadata.
A list of technical requirements for future analysis metadata systems addressing the needs of analysis at larger scales has been compiled, accounting for the diverse needs for metadata access in evolving analysis ecosystems, including practical and sociological concerns for individual analysers, sharing of information within experimental collaborations and the long-term preservation of analyses for reuse and reinterpretation.  These technical specifications follow the ``FAIR Guiding Principles for scientific data management and stewardship" allowing for Findable, Accessible, Interoperable and Reusable Metadata. 
While the discussions that led to this document primarily focused on experiments in High Energy Physics, the general principles and many of the specific challenges could also be applicable to experiments in other fields.
The list presented does not define particular solutions to the problems posed, but is rather intended to guide future R\&D on the concrete implementation of such systems.

\section{Acknowledgements}

This project is supported by funding from the European Research Council (ERC)
under the European Union's Horizon 2020 research and innovation programme: T. J. Khoo under
grant agreement 787331-HiggsSelfCoupling and N. Skidmore under grant agreement 852642-Beauty2Charm. S. Rappoccio is supported by the National Science Foundation under grant 2111229. 


\section{Declarations}
\subsection*{Data availability} Data sharing not applicable to this article as no datasets were generated or analysed during the current study.

\subsection*{Competing Interests}
On behalf of all authors, the corresponding authors state that there is no conflict of interest.

\bibliographystyle{unsrtdin}
\bibliography{main}


\end{document}